\documentclass[twoside,onecolumn,11pt]{article}
\usepackage{extsizes}
\usepackage[super,sort&compress,comma]{natbib} 
\usepackage[version=3]{mhchem}
\usepackage[left=3cm, right=3cm, top=3cm, bottom=3.0cm]{geometry}
\usepackage{doi}
\usepackage{balance}
\usepackage{amssymb}
\usepackage{mathptmx}
\usepackage{sectsty}
\usepackage{graphicx} 
\usepackage{lastpage}
\usepackage[format=plain,justification=justified,singlelinecheck=false,font={stretch=1.125,small,sf},labelfont=bf,labelsep=space]{caption}
\usepackage{subcaption}
\usepackage{float}
\usepackage{fancyhdr}
\usepackage{fnpos}
\usepackage[english]{babel}
\addto{\captionsenglish}{%
  
}
\usepackage[usenames,dvipsnames]{xcolor}
\usepackage{array}
\usepackage{droidsans}
\usepackage{charter}
\usepackage[T1]{fontenc}
\usepackage{orcidlink}
\usepackage{siunitx}
\usepackage{setspace}
\usepackage[compact]{titlesec}
\usepackage{hyperref}

\usepackage{epstopdf}
\usepackage{multicol}
\usepackage{tabularray}
\usepackage{booktabs}
\usepackage{tabularx}

\definecolor{cream}{RGB}{222,217,201}

\begin{document}
\pagestyle{fancy}
\thispagestyle{plain}
\fancypagestyle{plain}{
\renewcommand{\headrulewidth}{0pt}
}


\renewcommand{\thefootnote}{\fnsymbol{footnote}}
\renewcommand\footnoterule{\vspace*{1pt}%
\color{cream}\hrule width 3.5in height 0.4pt \color{black}\vspace*{5pt}} 
\setcounter{secnumdepth}{5}

\makeatletter 
\renewcommand\@biblabel[1]{#1}            
\renewcommand\@makefntext[1]%
{\noindent\makebox[0pt][r]{\@thefnmark\,}#1}
\makeatother 
\renewcommand{\figurename}{\small{Fig.}~}
\sectionfont{\sffamily\Large}
\subsectionfont{\normalsize}
\subsubsectionfont{\bf}
\setstretch{1.125} 
\setlength{\skip\footins}{0.8cm}
\setlength{\footnotesep}{0.25cm}
\setlength{\jot}{10pt}
\titlespacing*{\section}{0pt}{4pt}{4pt}
\titlespacing*{\subsection}{0pt}{15pt}{1pt}

\fancyfoot{}
\fancyfoot[LO,RE]{\vspace{-7.1pt}\includegraphics[height=9pt]{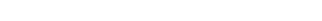}}
\fancyfoot[CO]{\vspace{-7.1pt}\hspace{13.2cm}\includegraphics{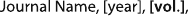}}
\fancyfoot[CE]{\vspace{-7.2pt}\hspace{-14.2cm}\includegraphics{head_foot/RF}}
\fancyfoot[RO]{\footnotesize{\sffamily{1--\pageref{LastPage} ~\textbar  \hspace{2pt}\thepage}}}
\fancyfoot[LE]{\footnotesize{\sffamily{\thepage~\textbar\hspace{3.45cm} 1--\pageref{LastPage}}}}
\fancyhead{}
\renewcommand{\headrulewidth}{0pt} 
\renewcommand{\footrulewidth}{0pt}
\setlength{\arrayrulewidth}{1pt}
\setlength{\columnsep}{6.5mm}
\setlength\bibsep{1pt}

\makeatletter 
\newlength{\figrulesep} 
\setlength{\figrulesep}{0.5\textfloatsep} 

\newcommand{\topfigrule}{\vspace*{-1pt}%
\noindent{\color{cream}\rule[-\figrulesep]{\columnwidth}{1.5pt}} }

\newcommand{\botfigrule}{\vspace*{-2pt}%
\noindent{\color{cream}\rule[\figrulesep]{\columnwidth}{1.5pt}} }

\newcommand{\dblfigrule}{\vspace*{-1pt}%
\noindent{\color{cream}\rule[-\figrulesep]{\textwidth}{1.5pt}} }

\makeatother

\twocolumn[
  \begin{@twocolumnfalse}
\vspace{1em}
\sffamily

\noindent\LARGE{\textbf{Optical materials discovery and design with federated databases and machine learning}} \\
\vspace{0.3cm}\\
{\noindent\large{Victor~Trinquet,\orcidlink{0009-0001-2188-4824}$^{\ast}$\textit{$^{a}$}
Matthew~L.~Evans,\orcidlink{0000-0002-1182-9098}$^{\ast}$\textit{$^{ab}$}
Cameron~J.~Hargreaves,\orcidlink{0000-0001-5509-9501}$^{a}$
Pierre-Paul~De~Breuck,\orcidlink{0000-0002-3173-2058}$^{\ddag}$$^{a}$
and Gian-Marco~Rignanese \orcidlink{0000-0002-1422-1205}$^{a}$}} \\
\vspace{0.3cm}\\
\noindent\normalsize{
Combinatorial and guided screening of materials space with density-functional theory and related approaches has provided a wealth of hypothetical inorganic materials, which are increasingly tabulated in open databases.
The OPTIMADE API is a standardised format for representing crystal structures, their measured and computed properties, and the methods for querying and filtering them from remote resources.
Currently, the OPTIMADE federation spans over 20 data providers, rendering over 30 million structures accessible in this way, many of which are novel and have only recently been suggested by machine learning-based approaches.
In this work, we outline our approach to non-exhaustively screen this dynamic trove of structures for the next-generation of optical materials. 
By applying MODNet, a neural network-based model for property prediction, within a combined active learning and high-throughput computation framework, we isolate particular structures and chemistries that should be most fruitful for further theoretical calculations and for experimental study as high-refractive-index materials.
By making explicit use of automated calculations, federated dataset curation and machine learning, and by releasing these publicly, the workflows presented here can be periodically re-assessed as new databases implement OPTIMADE, and new hypothetical materials are suggested.} \\


 \end{@twocolumnfalse} \vspace{0.6cm}

  ]

\renewcommand*\rmdefault{bch}\normalfont\upshape
\rmfamily
\section*{}
\vspace{-1cm}


\footnotetext{\textit{$^{a}$~UCLouvain, Institut de la Matiere Condens\'{e}e et des Nanosciences (IMCN), Chemin des \'{E}toiles 8, Louvain-la-Neuve 1348, Belgium}}
\footnotetext{\textit{$^{b}$~Matgenix SRL, 185 Rue Armand Bury, 6534 Goz\'{e}e, Belgium}}

\footnotetext{\dag~Electronic Supplementary Information (ESI) available: The structure and properties of the materials selected by the active learning loops for consideration with density-functional perturbation theory have been deposited to the Materials Cloud Archive (10.24435/materialscloud:5p-vq).}

\footnotetext{$\ast$~These authors contributed equally to this work.}
\footnotetext{$\ddag$~Present address: Ruhr-Universität Bochum, Universitätsstr. 150, 44801 Bochum, Germany}

\onecolumn  
\section{Introduction}


The advent of robust quantum mechanical calculations of material properties has expanded the opportunities for computational materials design.
It is now relatively commonplace for a single materials design study to consider vast swathes of the space of known inorganic compounds ($10^5$ entries), as curated in experimental~\cite{Bergerhoff1983, Zagorac2019} and computational~\cite{Jain2013, Kirklin2015} databases.
In recent years a large number of hypothetical materials have been suggested to be stable (for some definition) by data-driven and machine-learning approaches~\cite{Schmidt2023, Chen2021, Merchant2023, Zeni2024}.
These hypothetical materials can eventually make their way into curated databases, but often are released as static datasets that are hard to discover programmatically.
This new space is too large to study exhaustively, and the feasibility of said hypothetical materials requires significant attention~\cite{Cheetham2024}.
Data-driven screening methods must therefore be adapted to be able to rationale the most efficient allocation of experimental resources within this dynamic, growing, decentralised design space, especially when targeting specific material properties.


In this work, we devise a framework to search for materials with strong linear optical response, i.e., high-refractive-index (high-$n$) materials.
These are sought after for their application in waveguides, interference filters, mirrors, sensors, and anti-reflective coatings for solar cells~\cite{Naccarato2019,Higashihara2015Apr, Odom2015Jul, Chen2017Feb, Cheng2024, Liu2009Nov, senior2009optical}.
Nonlinear optical materials are typically used in signal processing, wavelength conversion, and quantum optics, but are more difficult to find~\cite{Chen1986Aug,Abudurusuli2021Mar, Dini2016Nov}.
As the linear optical response has been shown to be an indicator of non-linear (higher-order) optical response~\cite{Miller1964Jul, Meyer2024Apr}, high-$n$ semiconductors may be used as the starting point in the search for more exotic nonlinear optical properties.

For materials with the desired linear optical response, the electronic band gap, $E_g$, should be as wide as possible in order to guarantee transparency over the spectral range of interest, which will also increase the range of operating temperatures and voltages that a device can withstand~\cite{Watson2015Dec}.
However, the empirical trend is that the band gap is inversely proportional to the refractive index~\cite{Tripathy2015, Naccarato2019}.
These conflicting physical properties define a multi-objective Pareto front which must be optimised~\cite{Tamura2023}.
To overcome this challenge, this work combines machine learning property predictions with high-throughput screening of hypothetical materials databases to identify novel optical materials that best optimise this trade-off in different regions of the spectral range.



Previous high-throughput studies have used first-principles simulations\,\cite{Petousis2017, Naccarato2019} to predict the refractive index of thousands of semiconductors present in the Materials Project (MP)~\cite{Jain2013}.
The inverse trend linking $n$ and $E_g$ was observed in both studies, and modeled by \citet{Naccarato2019} with a two-variable parametric relationship derived from properties of the electronic structure.
Their main aim was to provide public datasets that could subsequently be screened to find promising high-$n$ compounds.
Here, we build on the dataset of 4,040 calculations provided in \citet{Naccarato2019} (henceforth N19).
The dataset and workflows of \citet{Petousis2017} have been integrated with the MP and now span $\sim$5,600 materials, here referred to as P17 (with some overlap with N19). 

As first-principles calculations are computationally demanding, optimising the set of candidate materials that are studied, without lowering the hit-rate of high-$n$ materials, is desirable.
To this end, two recent studies have used inexpensive machine-learned proxies for the refractive index to pre-screen compounds.

\citet{Carrico2024} trained a graph neural network on P17, to predict the refractive index of 55,000 materials contained in the Alexandria database~\cite{Schmidt2021, Schmidt2023}.
Materials whose predictions fell above the $(n, E_g)$ Pareto front of the N19 dataset were selected for validation.
The refractive index was then computed by first principles, and added to the training set for a second round of this process.
This approach successfully identified 2,431 compounds in the target region of the $n$-$E_g$ distribution (henceforth C24).

\citet{Riebesell2024} instead focused on high-$k$ dielectric properties, which additionally consider the ionic contribution to the dielectric tensor.
Using P17, they trained two deep ensembles of independent Wren models~\cite{Goodall2022}, to predict the ionic and electronic components of the dielectric tensor for all materials in the MP and the WBM dataset of \citet{Wang2021}.
An additional 100,000 trial structures were generated by chemical substitution for screening with the trained models.
All the candidate materials were then ranked according to a figure of merit (FoM) for miniming current leakage.
Finally, first-principles calculations were performed to compute the dielectric constant of the top-ranked candidates, yielding 2,691 hypothetical materials with computed dielectric tensors (referred to from now on as R24).

The present work follows a similar methodology to the above studies. 
The aim is to search for high-$n$ semiconductors by combining a machine-learned proxy for the refractive index and first-principles calculations as validation of promising candidates. 
This study differs by applying an extended active learning (AL) scheme, in which the process of "training-predicting-selecting-computing" is repeated multiple times, in contrast to the one or two iterations that \citet{Riebesell2024} and \citet{Carrico2024} respectively performed.
Our candidate pool generation also leverages the possibilities offered by OPTIMADE~\cite{Andersen2021} to simplify the querying of curated databases and their integration into the present workflow thanks to their standardised format.
The 30+ databases with OPTIMADE APIs provide a design space of over 27 million crystal structures which could be selected~\cite{Evans2024}.
Such a large, uncurated and decentralised set of structures presents unique opportunities and challenges for screening studies.

In this paper, we first introduce the computational workflow for computing the static refractive index, the implementation of our active learning procedure and the feedstock datasets used as the candidate pool.
As the result of several active learning campaigns, we computed the static refractive index of around 2,000 materials. 
Our selected materials were then screened against several additional criteria important for the realistic application of these materials.
Finally, we present the details of 21 of the most promising high-$n$ materials uncovered by our procedure and find that many have already been realised experimentally outside of the context of high-$n$ materials.

\section{Methods}


\subsection{Refractive index workflow}

In order to train a surrogate model via active learning, an ``oracle'' is required to define the ground truth property value for a given material so that the training dataset can grow over the AL iterations.
In this work, the electronic contribution to the dielectric tensor is computed in the framework of density-functional perturbation theory (DFPT)~\cite{Baroni1986May, Gonze1995Aug, Gajdos2006Jan}, as implemented in the Vienna Ab-Initio Simulation Package~\cite{Kresse1993Jan,Kresse1994May,Kresse1996Jul,Kresse1996Oct} (VASP version 6.3.0).
Following ~\citet{Petousis2017}, the static refractive index, $n_s$ or simply $n$, is given by:
\begin{equation}
    n_s = \sqrt{\varepsilon^{\infty}_\text{poly}} = \sqrt{\frac{\lambda_1 + \lambda_2 + \lambda_3}{3}},
\end{equation}
where $\varepsilon^{\infty}_\text{poly}$ is the average of $\lambda_i$, the eigenvalues of the electronic contribution to the dielectric tensor.

The exchange-correlation energy is modeled using the Perdew-Burke-Ernzerhof generalised-gradient approximation (GGA-PBE)~\cite{Perdew1996Oct, Perdew1997Feb} with Projector Augmented Wave pseudopotentials~\cite{Blochl1994Dec, Kresse1994Oct, Kresse1999Jan} (PBE\_64). The self-interaction energy is corrected with Hubbard $U$ values~\cite{Dudarev1998Jan} recommended by the MP~\cite{Jain2011Jun}. The plane wave energy cut-off was set to \SI{680}{\electronvolt} and the $k$-point grid was automatically generated with a density of 1,500 $k$-points per reciprocal atom.

These parameters balance efficiency with accuracy, keeping the relative error with respect to fully converged DFPT calculations below 10\%, as reported in \citet{Petousis2016Mar}.
This was confirmed on our dataset by computing $n_s$ with $k$-point grid densities of 1,500 and 3,000 on a representative set of 176 structures (with $\leq$ 5 atoms); the relative error on our high-$n$ biased dataset was observed to be much lower, less than 0.2\%, expanding the number of trials we could perform with a given compute budget.

The generation of inputs, the $k$-point grids, and the retrieval of outputs for the computation of the dielectric constants were automated with the \texttt{DielectricMaker} class of the \emph{atomate2} Python package~\cite{Ganose2024, Rosen2024}.
It is important to note that the crystal structures were not re-relaxed in this work, as all candidates were drawn from previously relaxed PBE calculations.

\subsection{Active learning}
\label{sec:AL}

In our AL loop, the MODNet architecture~\cite{DeBreuck2021} is used for the surrogate model.
MODNet consists of a relatively simple feed-forward neural network, with a relevance-redundancy-based feature selection algorithm applied to the broad set of featurisers implemented by the \emph{matminer} library~\cite{Ward2018}.
On the Matbench benchmark suite~\cite{Dunn2020}, MODNet has been shown to excel on datasets with fewer than 10k entries, and retains fast training and inference times on community grade hardware~\cite{DeBreuck2021a}.
By contrast, state-of-the-art graph models~\cite{Chen2019, Choudhary2021, Ruff2024} can require more data to attain sufficient accuracy, although transfer learning from more exhaustive datasets has shown promise~\cite{Hoffmann2023, Liu2022, Chen2021}. 
Other approaches targeted at small data, such as SISSO~\cite{Ouyang2018}, excel at interpretability over performance by constraining the inferred relationships via symbolic regression.

The \texttt{EnsembleMODNet} provides epistemic model uncertainty through ensemble variance which is an essential factor when carrying forward active learning on a predicted search space, as it enables a balance between exploration and exploitation.
Full exploration means that candidates with the highest uncertainty are chosen for oracle evaluation at each step of the AL loop, to be added to the training set in the following iteration.
Under this regime, the hope is that the ML model will improve whilst maximising data efficiency, as it is successively trained on the instances it was least certain about.
Full exploitation means that at each AL iteration, the candidates with the ``best'' predictions are chosen for  validation.
This means that computing resources are exclusively reserved for materials deemed promising by the model, but could overlook under-explored promising regions.

Our objective is to identify top candidates that exhibit both a high refractive index, $n$, and a high band gap, $E_g$.
Thus, we aim to discover new materials that extend the Pareto frontier in the $(n, E_g)$ space, with the most efficient use of computational resources.
To achieve this, we employ a strategy that combines cost-effective machine learning predictions with a more resource-intensive density-functional perturbation theory (DFPT) oracle.
A secondary aim is to devise a framework that can be iteratively reapplied in a manner akin to online learning, as new hypothetical materials are suggested in the literature and served through decentralised databases implementing OPTIMADE APIs~\cite{Evans2024}, with the aim of accelerating the suggestion of candidate materials for specific applications, and their subsequent experimental verification.

A pool-based active learning strategy is used, meaning that we sample from a discrete domain of fixed materials, $\mathcal{P}$, spanned by the publicly available OPTIMADE APIs that serve stability and band gap calculations.
At each learning cycle $i$, a MODNet ensemble is trained on training set $T_i$ in order to provide an estimate $p_i(n|x)\sim \mathcal{N}(\mu_{n,i},\sigma_{n,i})$ for a material $x$ with mean ensemble model prediction $\mu_{n,i}(x)$ and uncertainty $\sigma_{n,i}(x)$. From this, an upper bound to the refractive index, $n_U$, is computed:

\begin{equation}
  n_{U,i}(x) = \mu_{n,i}(x) +\lambda\sigma_{n,i}(x), \quad \lambda \geq 0,
\end{equation}
where $\lambda$ is a free parameter controlling the emphasis between exploration ($\lambda \to \infty$) and exploitation ($\lambda \to 0$), which can be tuned throughout the AL loop.
Following the methodology of~\citet{Riebesell2024}, a compromise is achieved by incorporating both the model prediction and its uncertainty in the acquisition function used at each step ($\lambda=1$).

Following the derivation of \citet{Naccarato2019}, we recast the multi-objective Pareto front optimisation as the maximisation of a semi-empirical effective frequency $\omega_\text{eff}$:

\begin{equation}
    \omega_\text{eff}(n,E_g) = \left(n^2 - 1\right)^{\frac{1}{3}}  \left( E_g + 6.74 - \frac{1.19}{E_g} \right)
    \label{eqn:w_eff}
\end{equation}

\noindent Although originally derived using only the direct band gap, here we approximate it with the fundamental gap for ease of comparison across multiple datasets. for The following acquisition function is then used to select materials for oracle evaluation (with e.g., DFPT):

\begin{equation}
  \alpha_i(x) = \omega_\text{eff}(n_{U,i}(x),E_g(x)).
  \label{eqn:w_eff_acquisition}
\end{equation}

\noindent The acquisition function $\alpha(x)$ is evaluated on the candidate pool of materials, $\mathcal{P}$. 
Given some user-defined compute budget, the materials that maximise the acquisition function are added to the next training phase: 
\begin{align}
  T_{i+1} &= T_i\,\cup\,\underset{x \in \mathcal{P}}{\mathrm{argmax}}\ \alpha_i(x)
\end{align}

\noindent A MODNet model is then trained on, $T_{i+1}$, and the loop can be repeated until sufficient accuracy, search coverage or materials performance is obtained.

Finally, it is worth pointing out the importance of correctly calibrating the uncertainty, $\sigma$, such that $p_i(n|x)$ matches the observed empirical distribution.
As shown in previous works, ensembles of MODNet models tend to give rather well-calibrated uncertainties~\cite{DeBreuck2022}.
This was confirmed on this dataset and displayed in Fig.~\ref{fig:calibration}, which shows the calibration curve and miscalibration area (MA) for our models on different AL iterations. 
One could reduce this area by a \emph{post hoc} rescaling of $\sigma$ that minimises the MA on a hold-out set.
Experiments show that a simple prefactor rescaling of $\sigma \to 1.3 \sigma$ is sufficient here.

\subsection{Training and candidate data}

The active learning loop was initialised with a training set of 4,040 density-functional perturbation theory (DFPT) calculations, $T_0$, containing the static refractive index and band gap from ~\citet{Naccarato2019} (N19).
For each entry in this dataset, the most recent structure was pulled from the Materials Project OPTIMADE API, when available. 
As 352 MP entries have been deprecated, 3,688 entries from the original N19 set remained.

A baseline ensemble MODNet model was trained to predict the refractive index on this dataset, using the \texttt{Matminer2024FastFeaturizer} preset implemented in MODNet v0.4.3, with relevance-redundancy criteria used to rank the computed features~\cite{DeBreuck2021}.
A genetic algorithm was used for hyperparameter optimisation, employed with 5-fold cross-validation to assess model performance~\cite{DeBreuck2022}.
The predictions of this baseline model are shown in \autoref{fig:naccarato-modnet}.
This baseline was used as the starting point for the AL loop.

The first set of trial structures were taken from the MP, which consists primarily of density-functional theory (DFT) relaxations of experimentally verified \cite{Bergerhoff1983, Zagorac2019} crystal structures (as well as many other computed properties).
Filtering the MP for phases with predicted band gaps greater than \SI{0.05}{\electronvolt} and distances from the MP convex hull (as computed by the latest mixed GGA+U/mGGA workflow~\cite{Kingsbury2022}) less than \SI{25}{\milli\electronvolt\per\text{atom}} results in 33,087 compounds, which we now refer to as the MP-33k set.
Any material with an existing refractive index calculation in the MP~\cite{Petousis2017} (around 5,000) was removed from the training or selection set, due to overlap with the N19 dataset.
Whilst these phases are all accessible through the MP's OPTIMADE API, the functionality to filter by band gap has not yet been exposed, so these phases were instead filtered using the MP's native API and then converted into the OPTIMADE structure format.

In addition to the MP, the AL loop was also able to explore a dataset constructed from the OPTIMADE API of the Alexandria PBE database~\cite{Schmidt2021, Schmidt2023}.
The OPTIMADE client implemented within the \href{https://github.com/Materials-Consortia/optimade-python-tools}{\emph{optimade-python-tools}} package~\cite{Evans2021} was used to filter for hypothetical compounds lying close to the Alexandria convex hull (again, \SI{25}{\milli\electronvolt\per\text{atom}}) and with a PBE-computed band gap greater than \SI{0.05}{\electronvolt}.
This approach provided an additional 104,860 structures to the design space which could be selected by the AL loop, henceforth Alexandria-105k.



\subsection{Implementation}
\label{sec:implementation}

The entire end-to-end workflow is implemented in a Python package named \texttt{re2fractive}, available on GitHub at \href{https://github.com/modl-uclouvain/re2fractive}{\texttt{modl-uclouvain/re2fractive}} under the MIT license.
Despite its name, this package has been designed to perform the entire active learning workflow on any physical property that MODNet can predict.

Each active learning run is controlled by an \texttt{re2fractive.Campaign} object, which is initialised with:
\begin{itemize}
    \item a programmatic description of an initial dataset that defines a set of properties, 
    \item a MODNet model type~\cite{DeBreuck2021, DeBreuck2021a} and the chosen preset of \emph{matminer} featurisers~\cite{Ward2018}, 
    \item a set of ``oracles'' for property values (e.g., \emph{atomate2} workflows~\cite{Ganose2024}), 
    \item an acquisition function comprised of the available properties,
    \item a set of logistical settings controlling the AL procedure and the calculation workflows (via Jobflow~\cite{Rosen2024} and jobflow-remote~\cite{JobflowRemote2024}).
\end{itemize}

The datasets can either be defined from a code pipeline which downloads, cleans and featurises a dataset directly (e.g., the N19 dataset~\cite{Naccarato2019}), or generating dynamic datasets with the latest data from a live OPTIMADE query.
A distinction must be made between the initial template dataset which contains the structure-property map (which can be empty), and the additional datasets used to define the design space, which are simply a source of trial structures.
These trial structures may come from other databases queried via OPTIMADE (or otherwise), but could be a pool generated by structure prediction, generative models, or other means.

At each step, the \texttt{Campaign} caches the datasets used, feature stores, and oracle results to disk, allowing the campaign to be restarted and extended from any point, with the performance at each checkpoint being plotted and recorded.

\subsection{Unsupervised embeddings}

Given the number of compounds in this study it is helpful to represent these in a condensed format, such as a plot, to observe the distribution of candidate materials in compositional space. 
The Element Movers Distance (ElMD)~\cite{Hargreaves2020} is an established metric of chemical similarity, which uses optimal transport to return a consistent measure of (dis)similarity between two given compositions that aligns with chemical intuition. 
An embedding of datasets with respect to this metric may be carried forward via kernel principal component analysis (PCA) by performing PCA on the ElMD distance matrix to give 2-dimensional coordinates for each material. 
It has been previously demonstrated that the Euclidean distances between resultant PCA embedded points are accurate realisations of the metric distances given by the ElMD \cite{Hargreaves2020}.

\section{Results}

\subsection{Learning procedure}

\begin{figure}[t]
     \centering
     \begin{subfigure}[t]{0.49\textwidth}
         \centering
         \includegraphics[width=\textwidth]{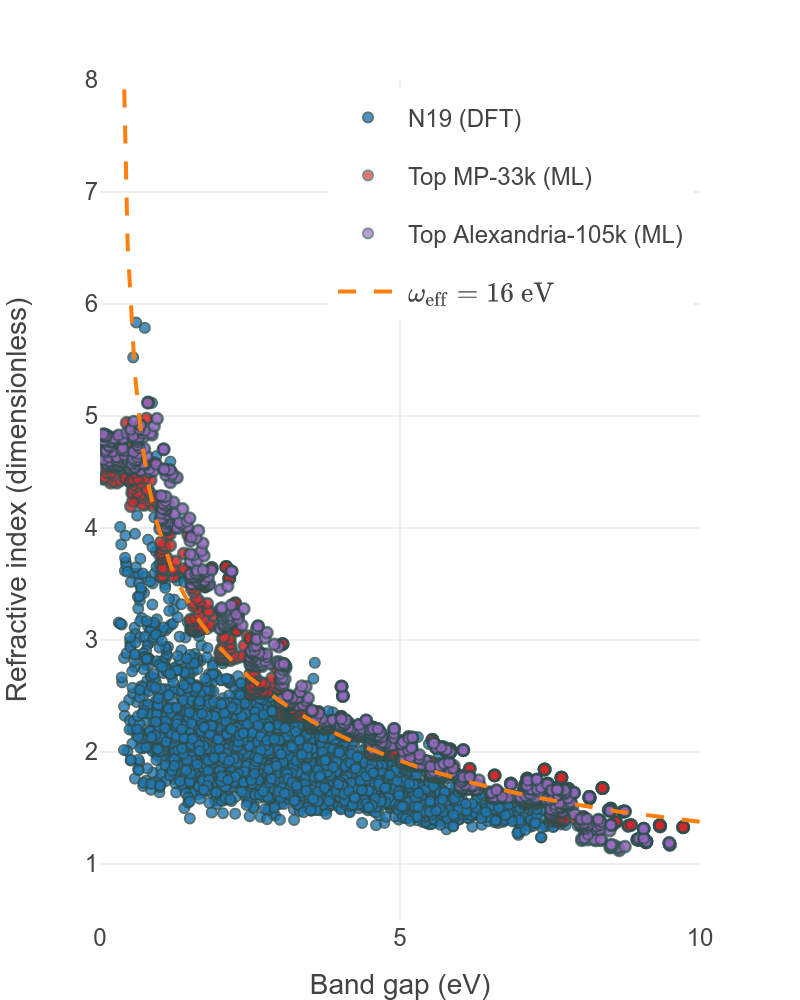}
         \caption{}
         \label{fig:naccarato-modnet}
     \end{subfigure}
     \hfill
     \begin{subfigure}[t]{0.49\textwidth}
         \centering
         \includegraphics[width=\textwidth]{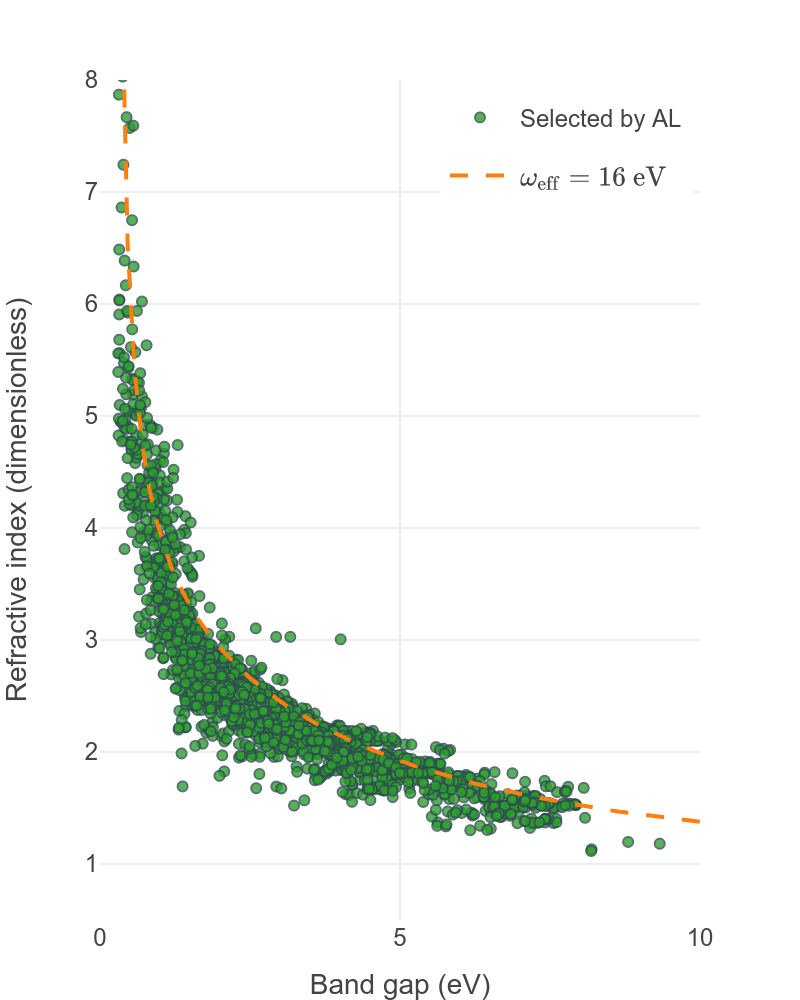}
         \caption{}
         \label{fig:al-final}
     \end{subfigure}
     
     \caption{The refractive index-band gap space spanned by the N19 dataset~\cite{Naccarato2019}. The left-hand panel, (a), shows the first iteration of model predictions on the ``design space'' datasets of MP-33k and Alexandria-105k. Only the top 50 predictions per \SI{0.5}{\electronvolt} band gap bin are displayed. The right-hand panel, (b), shows the DFT computed refractive index and band gap values for the compounds selected after 8 AL iterations by the $\omega_\text{eff}$ criteria described in \autoref{sec:AL}.
     \label{fig:design-space}}
\end{figure}

Several \emph{ad hoc} active learning runs were performed as described in \autoref{sec:AL} to select candidate structures on which to perform DFPT calculations.
At each selection step, an additional constraint was applied to have fewer than 100 sites in the unit cell, but structures exceeding this limit are still kept in the ML prediction pool.
All DFT calculations performed as part of these AL runs were saved in a global database such that future AL runs could make use of these results, without recomputing these properties.

We first quantified the advantage of active learning on the P17 dataset, compared to random and static learning strategies; the latter referred to as a single fitted model screening for candidates. 
Figures~\ref{fig:ALSL-fig1},~\ref{fig:ALSL-fig2} illustrate that a similar Top-100 score on P17 can be achieved with a lower oracle budget through active learning, thanks to iterative improvements made to the model during the search process. 
For further details, see Appendix~\ref{AL-SL}.

In our production runs, the model ensemble is trained on 80\% of the N19 dataset before making a selection, leaving a global holdout set for validation.
This initial ensemble achieved already an excellent accuracy on this holdout set, with an mean absolute error on the refractive index of 0.066 and mean absolute uncertainty of 0.038, consistent with the miscalibration area discussed previously.
The ensemble is then applied to the candidate pools spanned by the aforementioned MP-33k and Alexandria-105k datasets, as shown in \autoref{fig:naccarato-modnet}.
A number (initially 200, then 500) of the top candidates across these two sets were selected at each iteration $i$ for further DFPT computations, using the $\omega_\text{eff}(n_{U,i}(x), E_g(x))$ acquisition function introduced in \autoref{eqn:w_eff_acquisition}.
The AL loop was repeated 8 times to select the candidates comprising the final dataset shown in \autoref{fig:al-final}.
The N19 holdout error was not seen to decrease with further iterations, indicating that the selected materials were sampled from underexplored regions.
When evaluating the models \emph{post hoc} on the P17 and C24 sets (appropriately filtered and de-duplicated for any overlap with our selected candidates), we observe a much higher MAE around 0.2 and 0.4, respectively.

\begin{figure}[t]
    \centering
    \includegraphics[width=\textwidth]{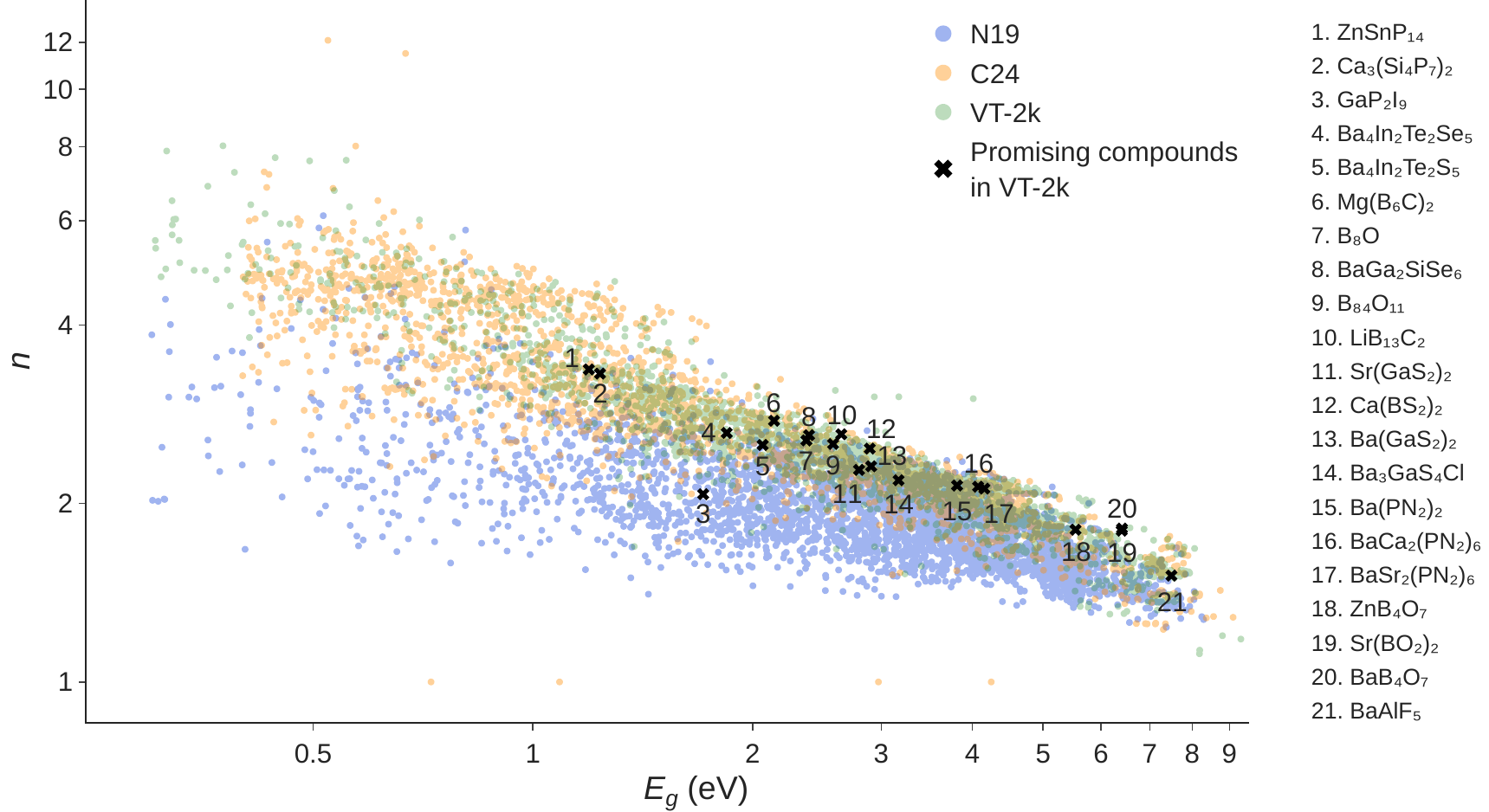}
    \caption{The $(n, E_g)$ space spanned by the N19~\cite{Naccarato2019} (blue), C24~\cite{Carrico2024} (orange) and VT-2k datasets (green).
    The numbered black crosses correspond to the materials that pass each of the filtering steps described in subsection~\ref{sec:filtering}, and listed in \autoref{tab:final_candidates}. \autoref{fig:screening-others} shows the result of applying the same filters to the C24 and R24 datasets.
    }
    \label{fig:vt2k-final}
\end{figure}



\label{sec:filtering}

The culmination of this work is a new, publicly available, dataset of static refractive index calculations for 2,413 materials, VT-2k. 
The dataset of structures, band gaps and refractive indices computed with the workflow described above is made available on the Materials Cloud Archive~\cite{VT-2k-MaterialsCloud} alongside the final model.
\autoref{fig:vt2k-final} depicts the full dataset superimposed on top of N19, the starting training set, and C24 as it was also designed to target high-$n$ materials. As hoped, both VT-2k and C24 are less disperse than N19 in the $(n, E_g)$ space. This implies that computational resources were used more efficiently in regard to the goal of identifying new high-$n$ compounds.
A similar representation with respect to P17 is shown in \autoref{fig:vt2k-P17}.

\begin{figure}[t]
\centering
\begin{subfigure}[b]{0.49\textwidth}
\includegraphics[width=\textwidth]{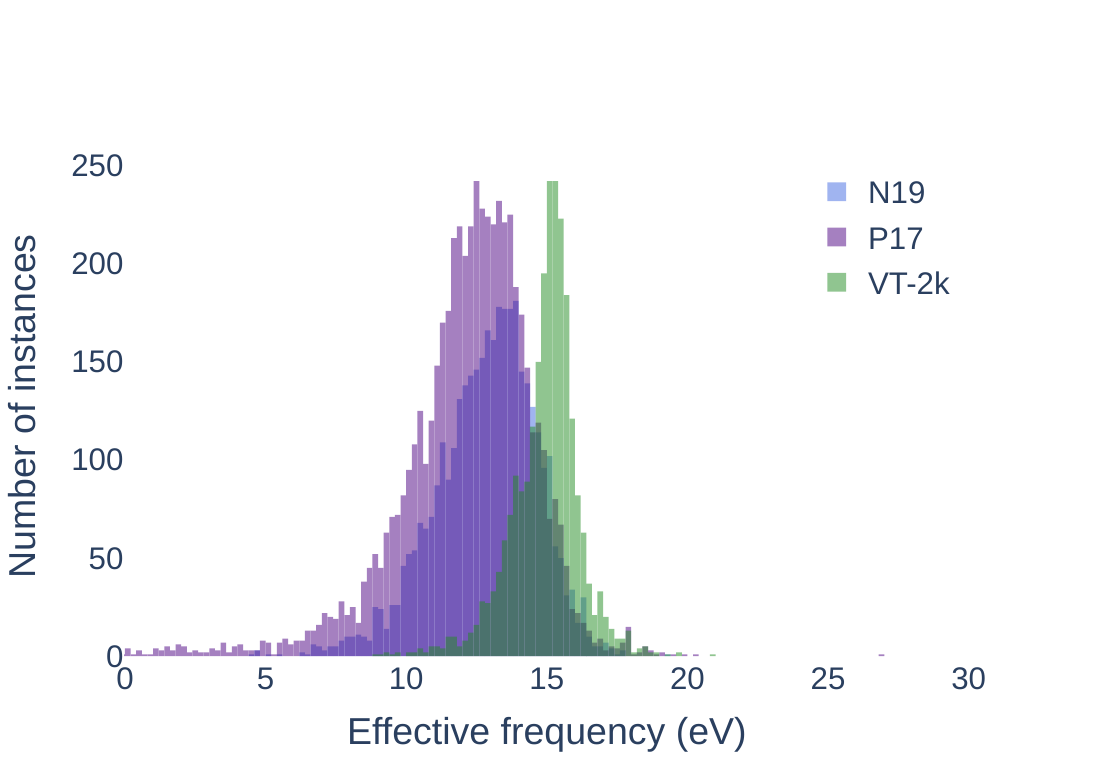}
\end{subfigure}
\hfill
\begin{subfigure}[b]{0.49\textwidth}
\includegraphics[width=\textwidth]{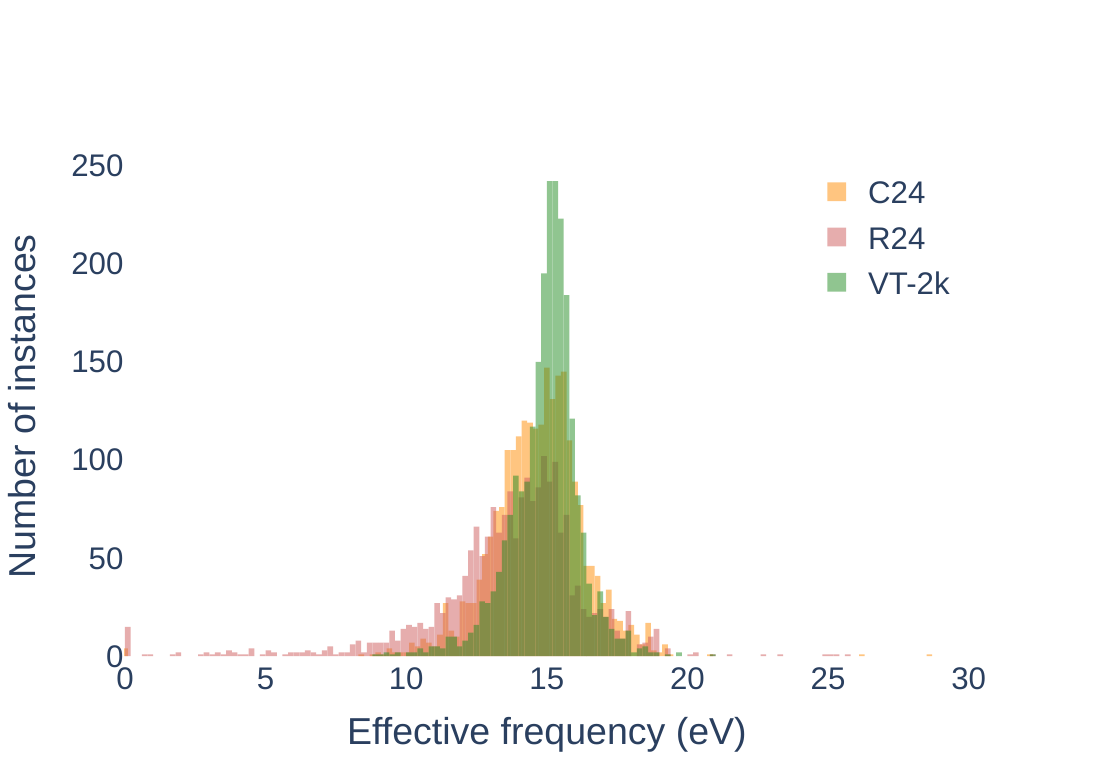}
\end{subfigure}
\caption{Comparison of the distributions of the effective frequency, $\omega_\text{eff}$, across the DFT datasets considered in this work.}
\label{fig:distrib_w_eff}
\end{figure}

\begin{figure}[t]
  \centering 
  \includegraphics[width=0.75\textwidth]{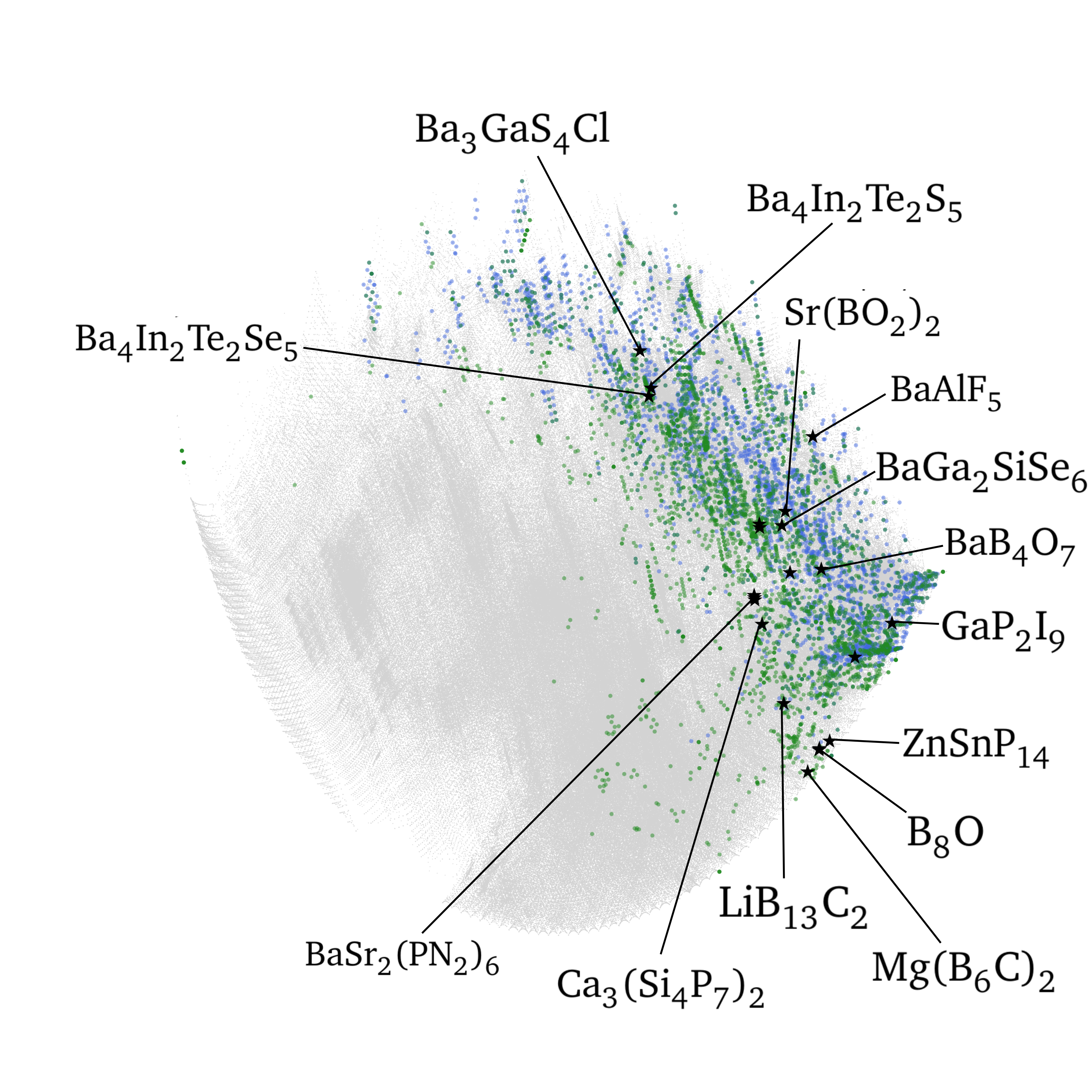} 
  \caption{An ElMD embedding of the N19~\cite{Naccarato2019} dataset (blue) and VT-2k dataset~\cite{VT-2k-MaterialsCloud} (green, this work) in the composition space spanned by 2M structures across the OPTIMADE consortium~\cite{Evans2024} (grey). The materials in Table 1 are overlaid in black stars, with a selection of these compounds labelled.} 
  \label{fig:elmd-vt2k}
\end{figure}

This sampling of high-$n$ materials is even more visible in \autoref{fig:distrib_w_eff} via the distribution of $\omega_\text{eff}$ (see \autoref{eqn:w_eff}) for the different datasets.
The upper panel shows that the pre-screening with a machine learned proxy has enabled to further push the Pareto front of high-$n$ materials when compared with the usual screening performed by \citet{Petousis2017} and \citet{Naccarato2019}. 
By comparison with C24, VT-2k displays a narrower distribution (lower panel), which may result from using a higher accuracy proxy to select an appropriate pool of trial structures for this target.
This can also be said for the comparison against R24 to some extent, although their goal was high-$k$ dielectrics instead of high refractive index compounds. \autoref{fig:elmd-vt2k} provides an ElMD embedding of this dataset, demonstrating the chemical diversity present. VT-2k is comprised of $\sim$43\% oxides (or at least, oxygen-containing compounds), with sulfur and selenium present in 16\% and 11\% of the entries respectively.
The dataset includes both simple and more complex structures, with $\sim$42\% of the computed entries having more than 20 atoms in their primitive unit cell.

\subsection{Filtering the resulting materials}

One of the goals of the present dataset is to offer a pool of materials for various screening procedures with emphasis on optical applications. Since the present work is not focused on a particular energy range, we illustrate a naive filtering by binning VT-2k with respect to the band gap (30 bins) and choosing the compound with the largest effective frequency in each bin. \autoref{tab:best_w_eff_candidates} lists the resulting candidates, which span a wide range of the electromagnetic spectrum and possess the best compromise between band gap and refractive index. However, out of the 30 structures, only 11 of them have been observed experimentally, and 9 of them contain lead, which is known for being toxic. This means that this method of selection lead to an oversampling of (potentially) unsynthesisable and toxic compounds, since VT-2k in aggregate consists of $\sim43$\% of experimentally observed materials and $\sim8$\% of lead-containing compositions. 
This simple procedure illustrates that, although the theoretical optimisation portion of this work treats the problem as a 2-objective optimisation, other factors must be considered when screening for realistic materials applications.
In order to identify the most promising high-$n$ materials for further investigation, the VT-2k compounds were screened by heuristics for the elemental sustainability, cost and availability, their capacity to be potentially synthesised in the future, and the figure of merit with respect to the targeted application. 

\begin{table}[t]
\scriptsize
\centering
\begin{tabularx}{\textwidth}{llcccccc}
\toprule
Identifier & Formula & $E_g$ (eV) & $n_s$ & $\omega_{eff}$ (eV) & Spacegroup & Exp. observed & MP ID \\
\midrule
mp-1028576 & \ce{Te8MoW3} & 4.011 & 3.006 & 20.940 & 156 & - & - \\
mp-1218922 & \ce{SnTe(PbS)4} & 0.555 & 7.592 & 19.781 & 160 & - & - \\
mp-571169 & \ce{InBi2Se4Br} & 1.295 & 4.741 & 19.778 & 12 & \checkmark & - \\
mp-1030335 & \ce{Te4MoW} & 3.172 & 3.028 & 19.206 & 156 & - & - \\
mp-1195735 & \ce{Te3As2} & 0.699 & 6.022 & 18.814 & 11 & \checkmark & - \\
mp-1028594 & \ce{Te4MoW} & 2.936 & 3.028 & 18.669 & 164 & - & - \\
agm002332658 & \ce{NbCoGe} & 1.074 & 4.726 & 18.596 & 216 & - & - \\
agm002233584 & \ce{PRhS} & 1.512 & 4.049 & 18.570 & 29 & - & - \\
mp-1212561 & \ce{Ge19(AsBr)4} & 1.425 & 4.106 & 18.415 & 218 & - & - \\
agm003202390 & \ce{YZn4Ir} & 0.956 & 4.879 & 18.294 & 216 & - & - \\
mp-1030319 & \ce{Te2Mo} & 2.596 & 3.104 & 18.215 & 164 & - & - \\
mp-1212555 & \ce{Ge19(PCl)4} & 1.650 & 3.752 & 18.067 & 218 & - & - \\
mp-4891 & \ce{HoAlO3} & 5.836 & 2.018 & 17.987 & 62 & \checkmark & - \\
agm003231828 & \ce{DyF3} & 8.062 & 1.680 & 17.897 & 165 & - & mp-1212889 \\
mp-4434 & \ce{TbAlO3} & 5.598 & 2.044 & 17.825 & 62 & \checkmark & - \\
mp-1035108 & \ce{LiMg14GaO16} & 6.873 & 1.810 & 17.682 & 47 & - & - \\
mp-21872 & \ce{PrLuO3} & 4.876 & 2.124 & 17.284 & 62 & \checkmark & - \\
mp-1212158 & \ce{Hf3GeO8} & 4.550 & 2.189 & 17.197 & 121 & - & - \\
agm002175849 & \ce{Hf3GeO8} & 4.352 & 2.207 & 16.989 & 215 & - & - \\
agm003214191 & \ce{Pr2PbS4} & 2.031 & 3.147 & 16.966 & 122 & - & mp-675638 \\
mp-33406 & \ce{La2PbSe4} & 1.828 & 3.290 & 16.957 & 122 & - & - \\
agm003209963 & \ce{PbICl} & 2.691 & 2.753 & 16.844 & 62 & \checkmark & mp-23053 \\
mp-36538 & \ce{La2PbS4} & 2.155 & 3.030 & 16.809 & 122 & - & - \\
mp-674363 & \ce{Zr3Sc4O12} & 3.947 & 2.237 & 16.492 & 1 & - & - \\
mp-1190186 & \ce{ZrO2} & 3.579 & 2.344 & 16.481 & 61 & \checkmark & - \\
mp-22997 & \ce{PbBrCl} & 3.438 & 2.385 & 16.456 & 62 & \checkmark & - \\
mp-1192803 & \ce{Lu2PbS4} & 2.302 & 2.822 & 16.279 & 62 & \checkmark & - \\
mp-669414 & \ce{HfPbO3} & 3.199 & 2.431 & 16.258 & 32 & \checkmark & - \\
agm002171412 & \ce{PrGaO3} & 3.816 & 2.209 & 16.097 & 167 & - & - \\
mp-1200793 & \ce{Ga4Ge(PbS3)4} & 2.367 & 2.740 & 16.065 & 114 & \checkmark & - \\
\bottomrule
\end{tabularx}
\caption{List of the 30 best materials based on the effective frequency of \autoref{eqn:w_eff}. Crystal structures with corresponding entries in the ICSD~\cite{Bergerhoff1983, Zagorac2019} are marked as ``experimentally observed'' and the MP ID is indicated for Alexandria materials matching an entry of the MP based on composition and spacegroup.} 
\label{tab:best_w_eff_candidates}
\end{table}

For the first criterion, the sustainability of each element can be characterised by the Herfindahl-Hirschman Indices (HHIs)~\cite{Herfindahl1950}.
These measure market concentration which, when applied to the geological and geopolitical availability of the element~\cite{Gaultois2013}, provides insight into possible future market abuse and subsequent price spikes of the materials.
Using the values from \citet{Gaultois2013} and following \citet{Kim2023}, we only consider elements with an HHI lower than 6000 for both production and reserves\footnote{Namely Li, B, C, N, O, F, Na, Mg, Al, Si, P, S, Cl, Ca, Zn, Ga, Ge, As, Se, Sr, Cd, In, Sn, Te, I, Ba, Hg, Pb}.
Compounds containing mercury (Hg), lead (Pb), cadmium (Cd), or arsenic (As) are also removed from this list due to their inherent toxicity.
It should be noted that HHIs are not static, and the values from \citet{Gaultois2013} are derived using geological surveying data from 2009-2011. As resources become more economically valuable, the number of geological surveys accelerate and new reserves are discovered, leading to oscillations in each of these indices.

To assess the potential synthesisability of a hypothetical material, typically, the main consideration is the predicted thermodynamic stability. By including only compounds with a predicted formation energy close to the global convex hull, i.e., whether the structure is at or close to the global minimum for that composition~\cite{Bartel2022}, materials which are unlikely to synthesise are excluded.
This is a reasonably strong, but imperfect, indicator of the synthesisability of a hypothetical material, as the experimental ground truth is heavily affected by kinetics, entropic considerations and the availability of precursor materials~\cite{Cheetham2024}.
In an attempt to model the difference between (zero temperature) DFT thermodynamic stability and realistically achievable synthesisability, recent works have created global reaction networks~\cite{McDermott2023} to find likely thermodynamic pathways, and machine learning models to directly predict relative synthesisability using semi-supervised or positively-unlabelled learning~\cite{Jang2020, Gu2022, Kim2024}.

In this work, we take the Composition Graph Neural Fingerprint (CGNF) atomic embedding provided by~\citet{Kim2024} (as trained on the MP~\cite{Jain2013} and OQMD~\cite{Kirklin2015}) to predict the probability of synthesisability of the hypothetical materials in VT-2k.
From these predictions, the third quartile of synthesisability predictions across VT-2k was used as a lower threshold for a material to be considered potentially synthesisable.
As a relative threshold is used, this approach is applicable to any dataset, and ensures that the materials in the upper tail of the synthesisability score distribution are selected.

Finally, to filter materials based on the figure of merit (FoM) for the intended application, an expressive metric that captures the specific utility of a material is required.
One could aim to minimise the current leakage ($\max{(\varepsilon_{poly}\cdot E_g)}$) or maximise the energy storage ($\max{(\sqrt{\varepsilon_{poly}}\cdot E_g)}$), however these FoMs would be more relevant in searching for high-$k$ dielectrics~\cite{Petousis2017, Riebesell2024}.
For high refractive index materials, an appropriate quantity to maximise is the effective frequency, $\omega_\text{eff}$, which was used during our AL selection process.
In order to generalise our method for future uses, the most promising materials can simply be defined as being those closest to the Pareto front of the properties to optimise.
In a 2-objective optimisation, these can be selected by binning the data over one objective (the band gap) and selecting all compounds above the third quartile of the second objective (the refractive index) for each bin.
This approach enforces the selection of promising materials over a broad spectral range, which would not necessarily be the case when using the effective frequency FoM alone.

As each filter uses a relative cut-off to screen the most promising materials, these operations are likely to be non-commutative, and the order the different filters are applied will influence the final choice of materials, which is illustrated in~\autoref{fig:final_screening}.
As presented, the selection of synthesisable and Pareto-optimal compounds leads to the exclusion of 75\% of the dataset.
The selection of low-HHI materials is even more drastic as 90\% of the dataset is discarded. 
At the final stage of each filter permutation, around 0.5\% of the dataset ($\sim$13 entries) is considered as promising and worth investigating.
The union of structures screened by each permutation of the filters comprises 21 materials; these are listed in \autoref{tab:final_candidates}.
It should be noted that this set already excludes any material covered in N19 or P17, which include most of the already known and stable optical materials.

\subsection{Promising materials}

\begin{table}[t]
\scriptsize
\centering
\begin{tabularx}{\textwidth}{lllcccccc}
\toprule
 & Identifier & Formula & $E_g$ (eV) & $n_s$ & $\omega_\text{eff}$ (eV) & Spacegroup & Exp. observed & MP ID \\
\midrule
1 & mp-1198754 & \ce{ZnSnP14} & 1.193 & 3.368 & 15.111 & 62 & \checkmark & - \\
2 & mp-1195188 & \ce{Ca3(Si4P7)2} & 1.236 & 3.314 & 15.101 & 14 & \checkmark & - \\
3 & mp-29817 & \ce{GaP2I9} & 1.711 & 2.074 & 11.550 & 61 & \checkmark & - \\
4 & mp-1193105 & \ce{Ba4In2Te2Se5} & 1.843 & 2.631 & 14.361 & 127 & \checkmark & - \\
5 & mp-1193666 & \ce{Ba4In2Te2S5} & 2.064 & 2.511 & 14.351 & 127 & \checkmark & - \\
6 & mp-568803 & \ce{Mg(B6C)2} & 2.140 & 2.759 & 15.621 & 74 & \checkmark & - \\
7 & mp-1228652 & \ce{B8O} & 2.370 & 2.556 & 15.225 & 6 & - & - \\
8 & mp-1227978 & \ce{BaGa2SiSe6} & 2.389 & 2.609 & 15.512 & 1 & - & - \\
9 & mp-758800 & \ce{B84O11} & 2.578 & 2.521 & 15.496 & 1 & - & - \\
10 & mp-655591 & \ce{LiB13C2} & 2.645 & 2.617 & 16.101 & 74 & \checkmark & - \\
11 & mp-14425 & \ce{Sr(GaS2)2} & 2.797 & 2.280 & 14.698 & 70 & \checkmark & - \\
12 & agm003284058 & \ce{Ca(BS2)2} & 2.894 & 2.476 & 15.905 & 205 & \checkmark & mp-30958 \\
13 & mp-849286 & \ce{Ba(GaS2)2} & 2.907 & 2.311 & 15.072 & 205 & \checkmark & - \\
14 & mp-1198976 & \ce{Ba3GaS4Cl} & 3.169 & 2.190 & 14.870 & 62 & \checkmark & - \\
15 & mp-2983 & \ce{Ba(PN2)2} & 3.811 & 2.145 & 15.697 & 205 & \checkmark & - \\
16 & mp-6404 & \ce{BaCa2(PN2)6} & 4.073 & 2.135 & 16.063 & 205 & \checkmark & - \\
17 & mp-567486 & \ce{BaSr2(PN2)6} & 4.154 & 2.120 & 16.097 & 205 & \checkmark & - \\
18 & agm003230381 & \ce{ZnB4O7} & 5.538 & 1.807 & 15.840 & 63 & \checkmark & mp-558690 \\
19 & agm003283778 & \ce{Sr(BO2)2} & 6.407 & 1.800 & 16.959 & 205 & \checkmark & mp-8878 \\
20 & agm003251671 & \ce{BaB4O7} & 6.412 & 1.815 & 17.101 & 62 & \checkmark & mp-556974 \\
21 & agm003272662 & \ce{BaAlF5} & 7.495 & 1.511 & 15.300 & 19 & \checkmark & mp-4376 \\
\bottomrule

\end{tabularx}
\caption{List of the most promising materials based on our screening with respect to the HHI, the synthesisability, and the quality as a high-$n$ compound, sorted by ascending $E_g$. Crystal structures with corresponding entries in the ICSD~\cite{Bergerhoff1983, Zagorac2019} are marked as ``experimentally observed'' and the MP ID is indicated for Alexandria materials matching an entry of the MP based on composition and spacegroup.} 
\label{tab:final_candidates}
\end{table}

Of the final set of 21 materials listed in \autoref{tab:final_candidates}, 18 have been previously observed experimentally.
By comparison, just 43\% of the broader VT-2k set have been experimentally observed.
This is an indicator that the crude ML-based synthesisability criteria is perhaps too biased towards known materials, or alternatively, that the hypothetical materials found predominantly in Alexandria-105k are too distant from the set of known materials to be considered plausible (without new directed synthesis campaigns).
The promising materials span a wide band gap range from \numrange[range-phrase=--]{1.2}{7.5}\,\unit{\electronvolt} whilst retaining a high $\omega_\text{eff} > 14.5\text{ eV}$.

In the low gap region (\numrange[range-phrase=--]{1.2}{1.8}\,\unit{\electronvolt}), \ce{ZnSnP14}--$Pnma$ (1)~\cite{Scholz1987}, \ce{Ca3(Si4P7)2}-$P2_1/c$ (2)~\cite{Zhang2015} and \ce{GaP2I9}-$Pbca$ (3)~\cite{Aubauer2001} are all previously isolated compounds with weakly bound structures and relatively large unit cells (50, 64 and 98 sites respectively), which may explain why they were excluded from previous high-throughput screenings. The reported colours of experimental powders in their initial synthesis reports~\cite{Scholz1987, Zhang2015, Aubauer2001} are consistent with the expected underestimated values predicted by DFT.

Two isostructural quaternary chalcogenides \ce{Ba4In2Te2$Q$5} ($Q=\text{S, Se}$) (4 and 5) crystallising in the $P4/mbm$ space group~\cite{Luo2013} are suggested with $\omega_\text{eff}$ around \SI{14}{\electronvolt}. They consist of layered \ce{[In2Te2$Q$4]^{6-}} covalent units, and UV-vis measurements of their respective band gaps (\SI{2.3}{\electronvolt} and \SI{1.8}{\electronvolt}) are broadly consistent with calculations, indicating that the computed refractive indices of $\sim$2.5 are feasible. The similar hypothetical quarternary barium chalcogenide \ce{BaGa2SiSe6}-$P1$ (8) (hull distance \SI{14}{\milli\electronvolt\per\text{atom}}) is also predicted to have a similar refractive index in this band gap range, though has not yet been isolated and is instead expected to decompose to the simpler \ce{Ba(GaSe2)2} phase (and \ce{SiSe2}) even without further entropic considerations.

The \numrange[range-phrase=--]{2.0}{2.6} \unit{\electronvolt} range is populated by boron-based materials, from the weakly bound \ce{B12} icosehedra in \ce{Mg(B6C)2}-$Imma$ (6)~\cite{Adasch2007}, and \ce{LiB13C2}-$Imma$ (10)~\cite{Vojteer2006} have both been previously isolated and, in the case of \ce{LiB13C2} with an excellent $\omega_\text{eff}$ value of \SI{16.1}{\electronvolt}, has been observed to be colourless.
Two hypothetical boron suboxides \ce{B8O}-$Pm$ (7) and \ce{B84O11}-$P1$ (9), derived from \ce{B6O}-$R\bar{3}m$~\cite{Perevislov2022}, have previously shown promise for their mechanical properties as composites, but are hard to isolate.

Between \SI{2.7}{\electronvolt} and \SI{3.2}{\electronvolt}, the selected materials are dominated by relatively well-known sulphide materials, including the thiogallates \ce{Sr(GaS2)2}-$Fddd$ (11), \ce{Ba(GaS2)2}-$Pa\bar{3}$ (13), \ce{Ba3GaS4Cl}-$Pnma$ (14) and the related derivative \ce{Ca(BS2)2}-$Pa\bar{3}$ (12)~\cite{Sasaki2003} (seemingly only accessible via high-pressure synthesis).
These compounds have previously been studied as phosphor materials~\cite{Peters1972} for their response under ultraviolet light.

In the wide gap range above \SI{3.8}{\electronvolt}, three barium nitrodophosphates \ce{Ba(PN2)2}-$Pa\bar{3}$ (15), \ce{BaCa2(PN2)6}-$Pa\bar{3}$ (16) and \ce{BaSr2(PN2)6}-$Pa\bar{3}$ (17)~\cite{Karau2006} have all been previously observed with relatively large unit cells containing 84 atoms. These are the final three materials considered that would be transparent in the visible range, whilst still retaining a refractive index greater than 2.

Finally, three very wide-gap oxides \ce{ZnB4O7}-$Cmcm$ (18), \ce{Sr(BO2)2} (19) and \ce{BaB4O7} (20) are suggested with excellent $\omega_\text{eff}$ values, driven by their band gaps which exceed \SI{5.5}{\electronvolt}. These materials have only ever been isolated at high pressures~\cite{Huppertz2003, Ross1991, Knyrim2009} but possess relatively low predicted hull distances (below \SI{25}{\milli\electronvolt\per\text{atom}}.
Beyond these, the \ce{BaAlF5}-$P2_1 2_1 2_1$ (21)~\cite{Domesle1982} phase with a band gap of \SI{7.5}{\electronvolt} has the widest gap considered here, and whilst its refractive index is just 1.5, this still corresponds to a competitive $\omega_\text{eff}$ of \SI{15.3}{\electronvolt}.

As previously mentioned, these results must be caveated by the fact that GGA-based DFT typically underestimates band gaps in an uncontrolled and unsystematic way; this will also be reflected in the overestimation of the refractive index.
However, we believe this set of materials comprise a promising set to consider for linear (and potentially nonlinear) optical applications, which may motivate further attempts to synthesise those that have not yet been isolated.

\section{Conclusions}


In recent years, the field of materials informatics has evolved from curating databases of calculations on known materials ($\sim10^5$)~\cite{Jain2013}, to directly predicting crystal structures at a given composition ($\sim10^3$)~\cite{Harper2020a, Goodall2022}, to using machine learning approaches to suggest stable hypothetical materials ($\sim 10^6$)~\cite{Chen2022, Merchant2023}, increasingly targeted towards particular properties~\cite{Alverson2022, hargreaves2023database, Zeni2024}.
As curated materials datasets grow and new hypothetical materials are suggested, it is imperative that we are able to effectively screen these compounds to guide experimental work towards the most fruitful areas.
We stress the importance of a decentralised approach to data sharing following the model of OPTIMADE, that complements the work done in curating high-quality databases, as this enables materials informatics studies to screen our best and latest guess of the global phase diagram in a machine-actionable manner.
With this in mind, there are three directions in which the present work will be extended.

Firstly, the procedure for screening high-$n$ materials will be applied to additional datasets as they become available with OPTIMADE APIs automatically~\cite{Andersen2021}. 
This takes advantage of the decentralised nature of the OPTIMADE consortia to encompass not only the growing number of structures available in curated experimental databases such as the ICSD~\cite{Bergerhoff1983, Zagorac2019} (and their corresponding calculations in the MP~\cite{Jain2013}), but also hypothetical structures provided in small datasets~\cite{Evans2020}, ensuring that each screening step is given a broad exposure to the landscape of hypothetical materials. 
This work will be used in part to motivate the ongoing development of the OPTIMADE stability namespace~\cite{Evans2024}, which will enable data unification across the providers that serve crystal structures with corresponding thermodynamics calculations. 
This would enable our AL campaign to run indefinitely and autonomously, adding to the dataset of validated materials based on a simple multi-provider OPTIMADE query.\footnote{Such an OPTIMADE query could be $ \texttt{\_stability\_hull\_distance <= 0.05\ AND\ last\_modified >= "2024-09-10-00:00"}$} 
All of the results generated by this online procedure will be made available themselves via an OPTIMADE API, to allow other human and machine agents to make use of our predictions.
Additionally, generative approaches will be investigated that make use of the final AL-trained model as a discriminator for high-$n$ materials.
These generative trials can then be used to iterate much more rapidly, without needing to wait for the results of external materials discovery campaigns to be made available.

Secondly, the best-performing materials screened in this investigation will be examined with more scrutiny before it is possible to propose them for costly experimental synthesis and characterisation.
This will include incorporating calculations of band gaps at higher levels of theory, which may then re-enter future AL campaigns via multi-fidelity learning~\cite{DeBreuck2022}.
The filtering procedure described in \autoref{sec:filtering} may be further refined (e.g., via improvements to the synthesisability filter, or by relaxing some of the constraints on HHI based on the latest data).

Finally, the resulting models and datasets created here for high-$n$ materials will be used to initialise a campaign searching for materials with more exotic physical properties, such as nonlinear optical response.
We believe the \texttt{re2fractive} AL approach is particularly well suited for this case, where there are significantly fewer calculations available, and each individual calculation (or oracle evaluation) requires significantly more computational resources to execute.
In this sense, the entire AL campaign performed in this work will form part of the selection criteria of a broader campaign, and the \texttt{re2fractive} framework outlined in \autoref{sec:implementation} will be expanded to accommodate this stacked approach.



\clearpage

\section*{Author Contributions}

\textbf{V.T.}: Conceptualisation, Methodology, Software, Validation, Investigation, Data curation, Writing - Original Draft, Writing - Review and Editing, Visualisation.
\textbf{M.L.E.}: Conceptualisation, Methodology, Software, Validation, Investigation, Data curation, Writing - Original Draft, Writing - Review and Editing, Visualisation.
\textbf{C.J.H.}: Conceptualisation, Writing - Original Draft, Writing - Review and Editing, Visualisation.
\textbf{P-P.D.B.}: Conceptualisation, Methodology, Validation, Writing - Original Draft.
\textbf{G-M.R.}: Conceptualisation, Resources, Writing - Review and Editing, Supervision, Funding acquisition.

\section*{Conflicts of interest}
G.-M.R. is a shareholder and Chief Innovation Officer of Matgenix SRL.

\section*{Acknowledgements}
Computational resources have been provided by the supercomputing facilities of the Université catholique de Louvain (CISM/UCL) and the Consortium des Équipements de Calcul Intensif en Fédération Wallonie Bruxelles (CÉCI) funded by the Fond de la Recherche Scientifique de Belgique (F.R.S.-FNRS) under convention 2.5020.11 and by the Walloon Region. The present research benefited from computational resources made available on Lucia, the Tier-1 supercomputer of the Walloon Region, infrastructure funded by the Walloon Region under the grant agreement n°1910247. V.T. acknowledges the support from the FRS-FNRS through a FRIA Grant. M.L.E. thanks the BEWARE scheme of the Wallonia-Brussels Federation for funding under the European Commission's Marie Curie-Skłodowska Action (COFUND 847587). C.J.H. thanks the FRS-FNRS for their support as part of the Fish4Diet project. 

\balance

\bibliography{re2fractive} 
\bibliographystyle{unsrtnat}

\appendix
\renewcommand\thefigure{\thesection.\arabic{figure}}
\setcounter{figure}{0}
\section{Appendix}
\subsection{Supplementary figures}
\begin{figure}[H]
  \centering 
  \includegraphics[width=0.65\linewidth]{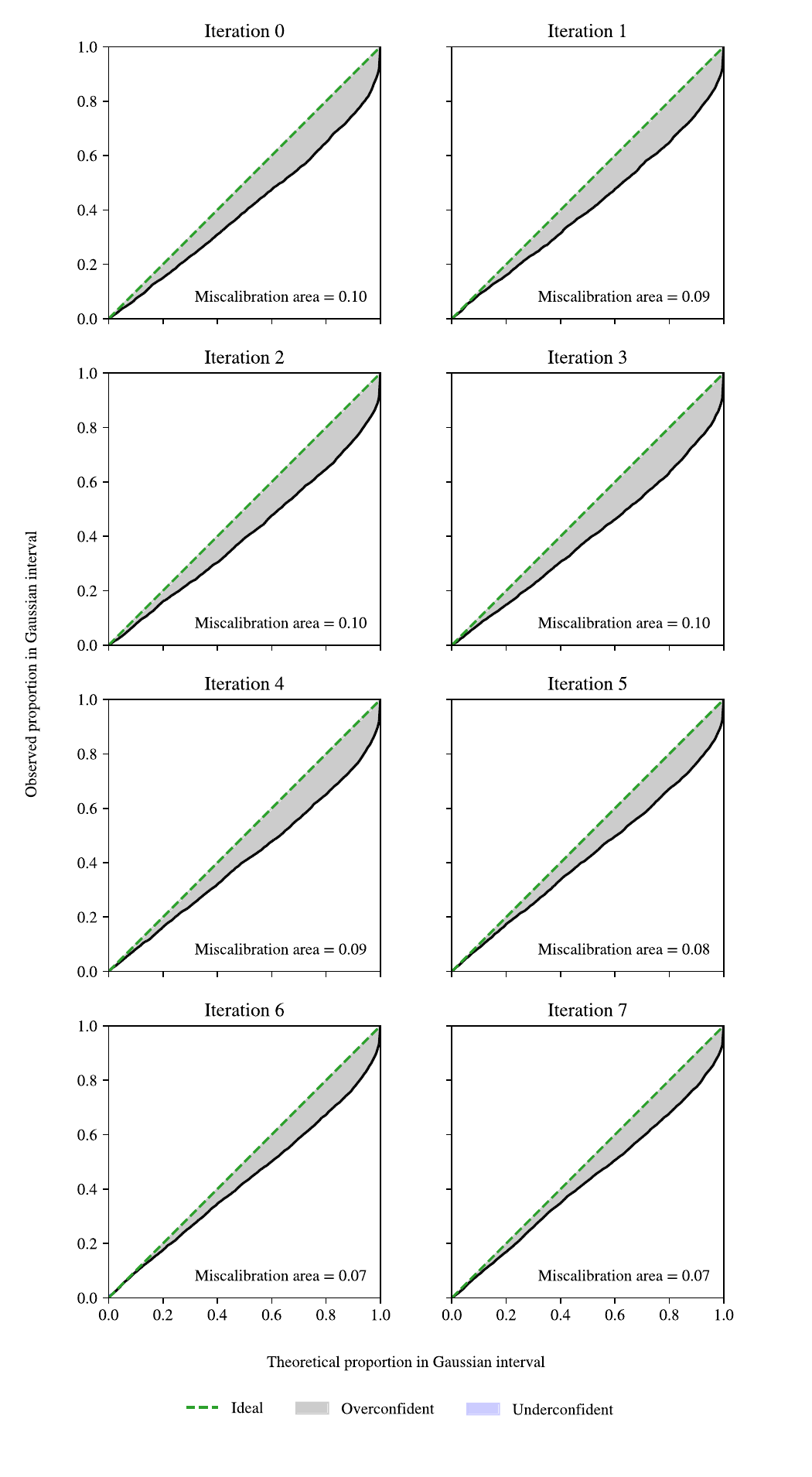} 
  \caption{Calibration curves for the MODNet model used in the different iterations of the active learning campaign} 
  \label{fig:calibration}
\end{figure}

\begin{figure}[h]
\centering
\begin{subfigure}[b]{\textwidth}
\centering
\includegraphics[width=\textwidth]{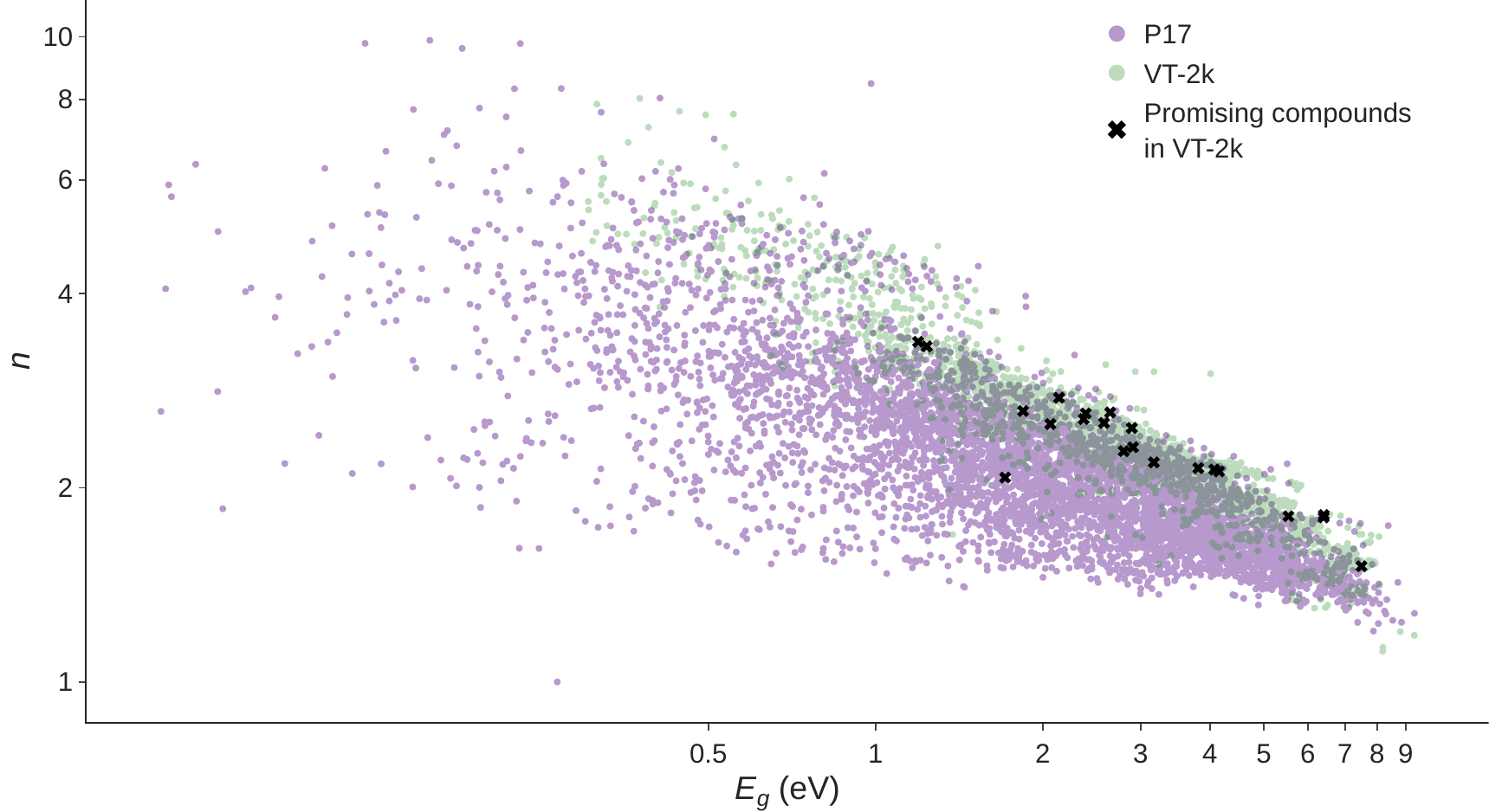}
\end{subfigure}
\caption{The $(n, E_g)$ space spanned by the P17~\cite{Petousis2017} (purple), and VT-2k datasets (green).
    The black crosses correspond to the materials that pass each of the filtering steps described in \autoref{sec:filtering}. They are also listed in \autoref{tab:final_candidates}.}
\label{fig:vt2k-P17}
\end{figure}

\begin{figure}[h]
\centering
\begin{subfigure}[b]{\textwidth}
\centering
\includegraphics[width=\textwidth]{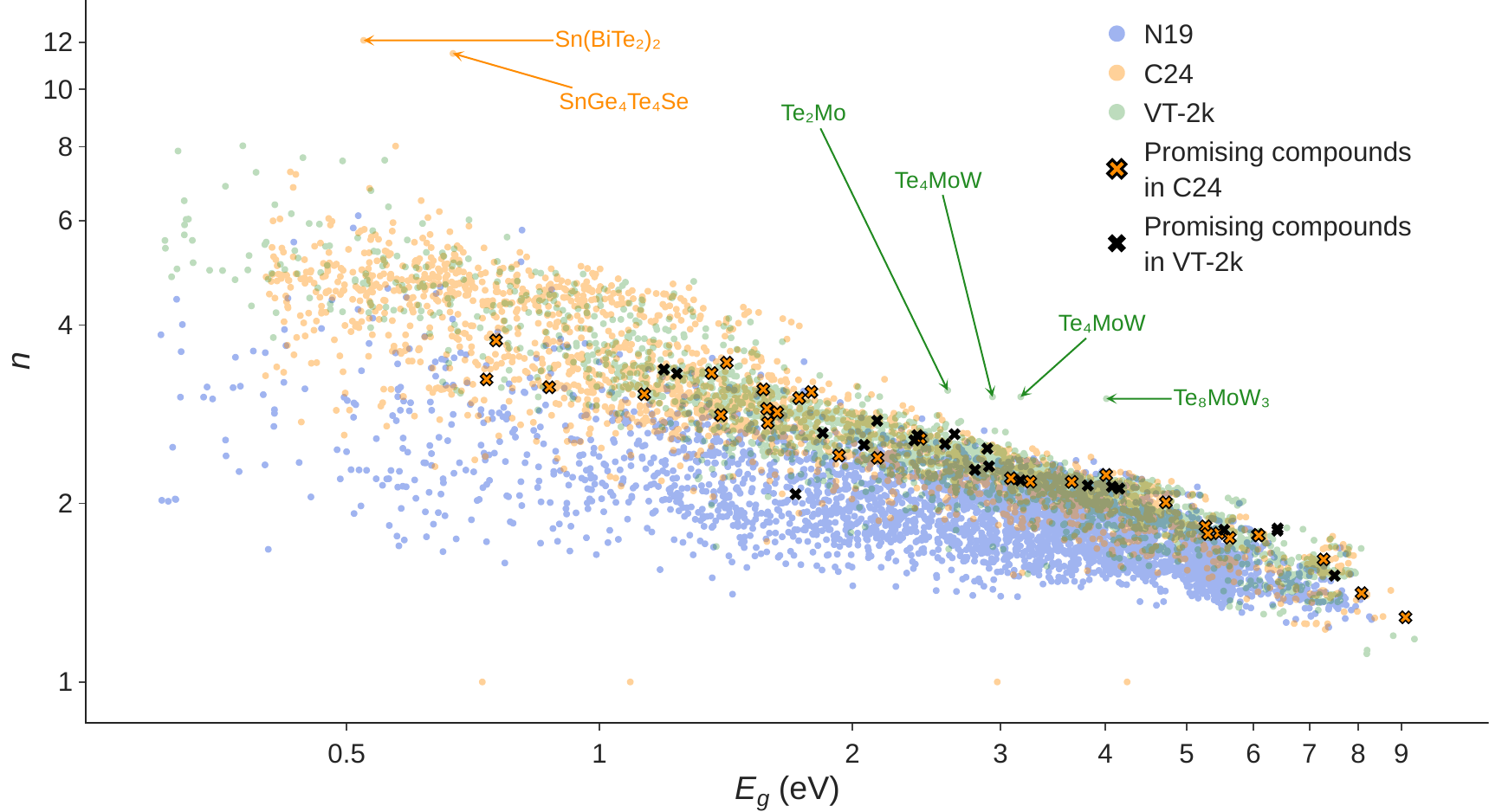}
\end{subfigure}
\begin{subfigure}[b]{\textwidth}
\centering
\includegraphics[width=\textwidth]{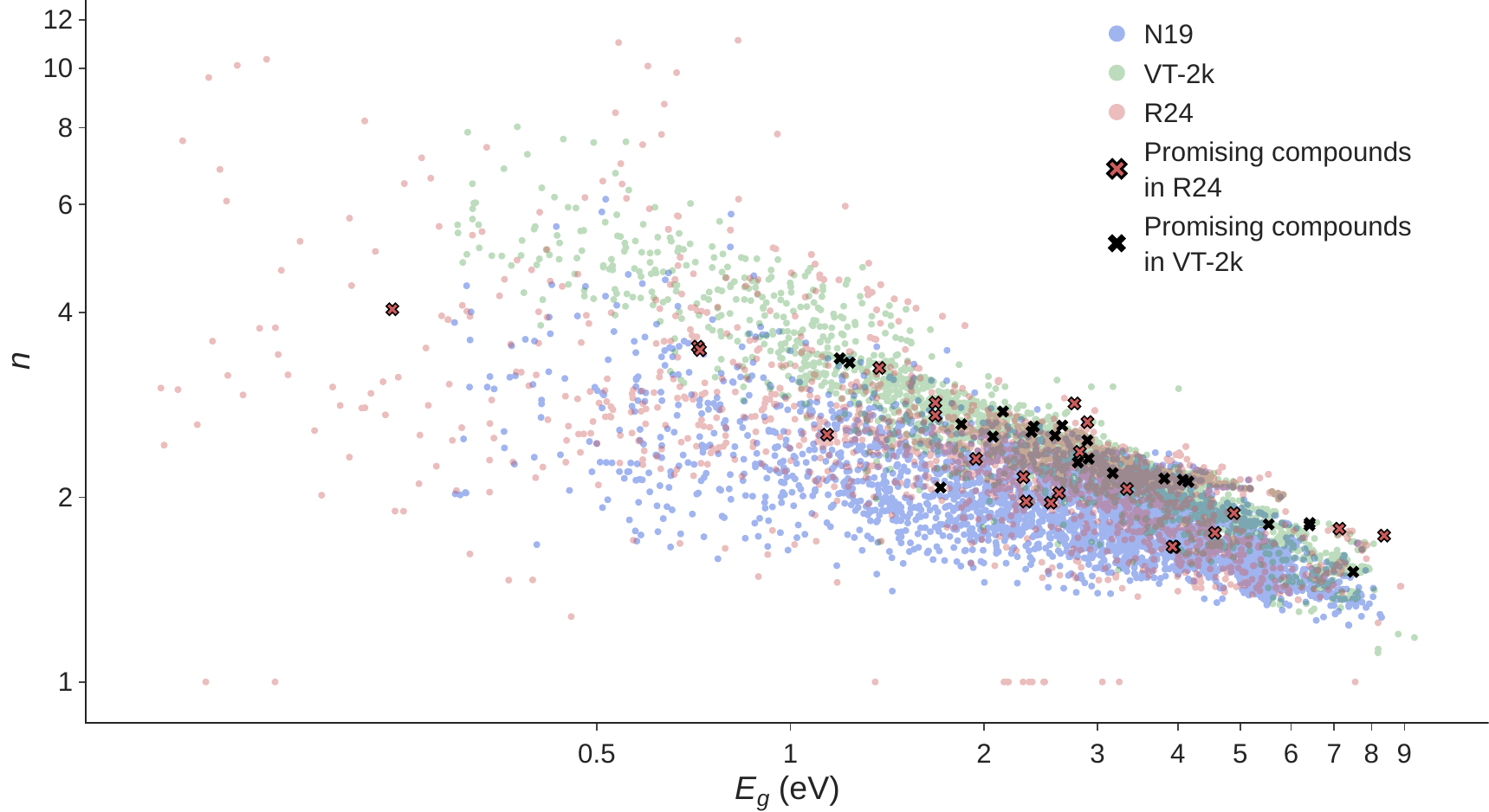}
\end{subfigure}
\caption{The filters described in \autoref{sec:filtering} applied to the external datasets of C24~\cite{Carrico2024} (orange crosses, upper panel) and R24~\cite{Riebesell2024} (red crosses, lower panel), in comparison with the filtered materials from VT-2k (black crosses). Exceptional outlier materials (Te-W-Mo and Sn-Te containing compounds) from each dataset are labelled explicitly.}
\label{fig:screening-others}
\end{figure}

\begin{figure}[h]
  \centering 
  \includegraphics[width=0.8\textwidth]{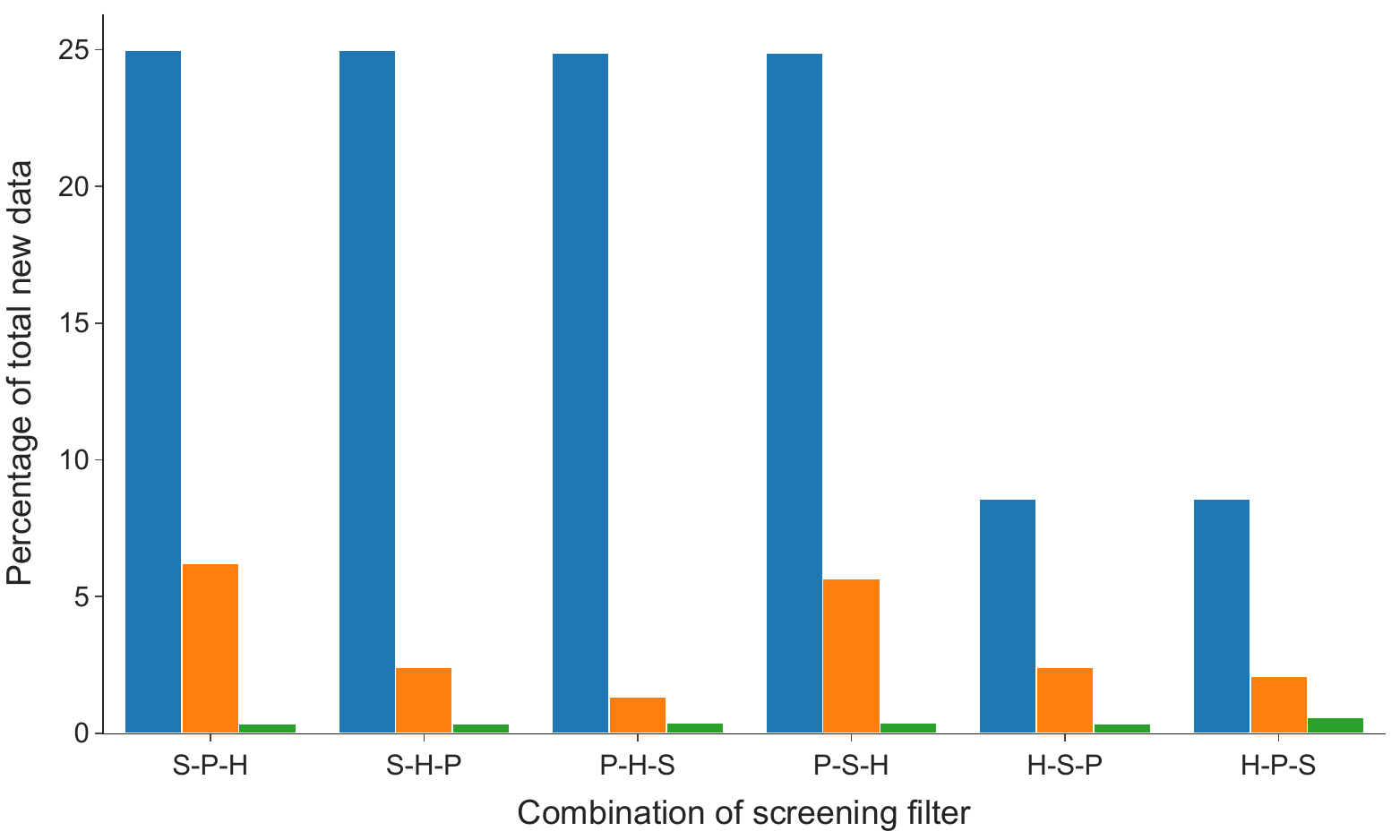} 
  \caption{Effect of the order of the filters on the final screening of promising high-$n$ materials. The y-axis shows the relative quantity of selected materials with respect to the present dataset (2,413 entries).
  The different orders of filters are situated along the horizontal axis where 'S' stands for the criterion on synthesisability, 'P' for the restriction to the Pareto neighbourhood, and 'H' for the exclusion of high HHI elements.} 
  \label{fig:final_screening}
\end{figure}

\clearpage

\subsection{Active versus static learning}
\label{AL-SL}

The advantage of active learning with respect to other common search methods is demonstrated using the P17 dataset. Initially, a subset of 1,000 labeled samples from P17 is designated as the starting set $\mathcal{L}$ (simulating a known dataset), while the remaining $\sim$ 5,000 data points form the candidate pool $\mathcal{P}$ and remain unlabeled (simulating known structures with unknown characteristics).
Different strategies are then employed to query $b$ points (as defined by the budget, up to 1,000) from $\mathcal{P}$.
These points are subsequently labeled using the P17 dataset (simulating an oracle).
The quality of the selected points is quantified by computing the top-$k$ score over $\omega_{\text{eff}}$:

\begin{equation*}
    \text{top}_k(\omega_{\text{eff}}) = \frac{1}{k} \sum_{i=1}^{k} \omega_{\text{eff}, (i)},
\end{equation*}

\noindent where $\omega_{\text{eff}, (i)}$ represents the $i$-th highest score among the newly labeled samples. This metric is calculated over the $k$ samples with the highest $\omega_{\text{eff}}$, where $k$ is an adjustable parameter.

\begin{figure}[ht]
  \centering 
  \includegraphics[width=0.85\linewidth]{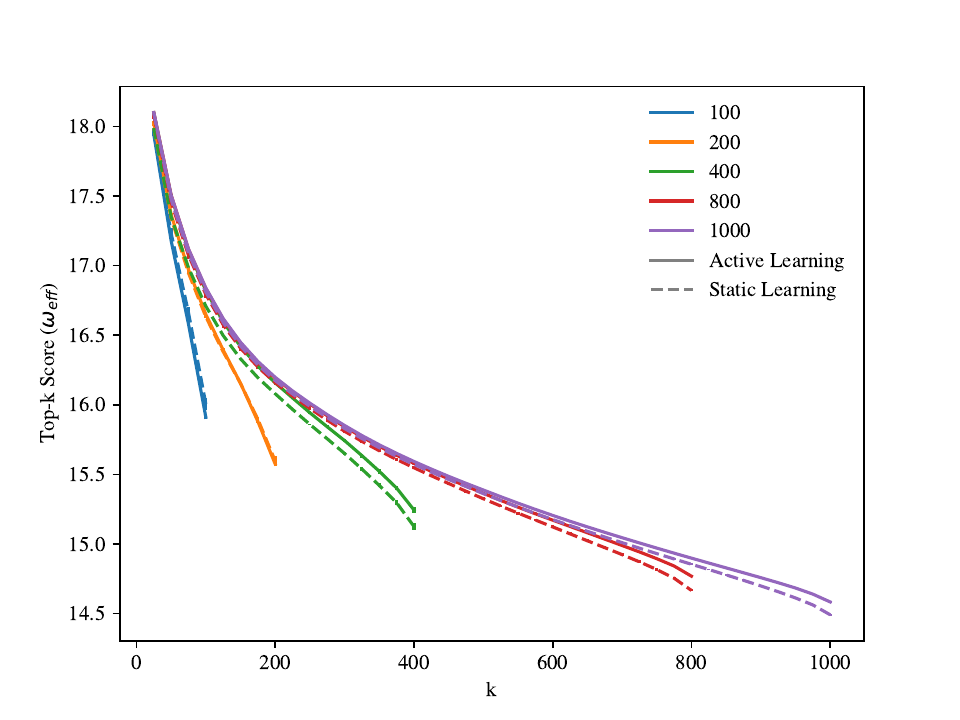} 
    \caption{Top-$k$ score as a function of $k$ for different search budgets, comparing AL (Active Learning) and SL (Static Learning) algorithms.}

  \label{fig:ALSL-fig1}
\end{figure}

As a baseline strategy, one might randomly select points from $ \mathcal{P} $. The P17 dataset presents average values of $ n = 2.29 \pm 0.86 $, $ E_g = 2.44 \pm 1.67 $, and $ \omega_{\text{eff}} = 11.91 \pm 3.61 $ ($\pm$ denoting the standard deviation). For a random search baseline, the expected value of the top-$k$ score, calculated over the highest $ k $ scores from randomly selected points, would approximate the mean of $ \omega_{\text{eff}} $, i.e., 11.91. The variance of this score would decrease with increasing $ k $, scaling approximately as $ 1/k^2 $.

A more commonly used approach is to train a surrogate machine learning (ML) model on $ \mathcal{L} $, and then use it to screen $ \mathcal{P} $ for promising candidates, selecting the $ b $ candidates with the highest predicted $ \omega_{\text{eff}} $. We refer to this method as static learning (SL). In contrast, active learning (AL) iteratively updates its ML model while exploring $ \mathcal{P} $. Both strategies are allowed to query $ b $ new samples from $ \mathcal{P} $, as defined by the budget.

Figure~\ref{fig:ALSL-fig1} represents the top-$k$ score as a function of $k$, for different budget sizes. It is observed that AL systematically outperforms SL. Although the advantage is small, it is noteworthy that with a budget of 800, AL achieves a similar or better top-$k$ score than SL with a budget of 1000, across almost the entire range of $k$. This represents a non-negligible saving of resources of 20\%.

\begin{figure}[ht]
  \centering 
  \includegraphics[width=0.85\linewidth]{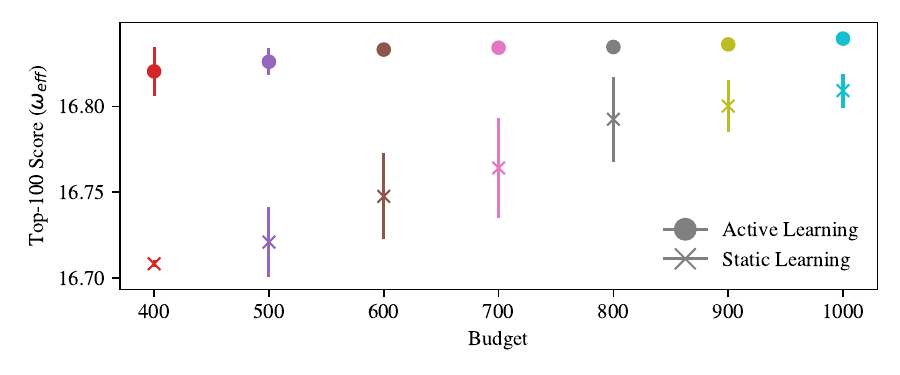} 
    \caption{Top-100 score as a function of iterations (i.e., search budget) for AL (Active Learning) and SL (Static Learning) search algorithms. Error bars represent the standard deviation over 5 repeated experiments.}
  \label{fig:ALSL-fig2}
\end{figure}

Figure~\ref{fig:ALSL-fig2} displays the Top-100 score for various budgets, while also depicting the observed standard deviation over repeated experiments. A slight yet consistent advantage is observed for active learning, with a notably smaller variance among selected candidates. Notably, a budget of 1000 is necessary for static learning to achieve a Top-100 score comparable to that reached by active learning with just a budget of 400.

\begin{figure}[ht]
  \centering 
  \includegraphics[width=0.85\linewidth]{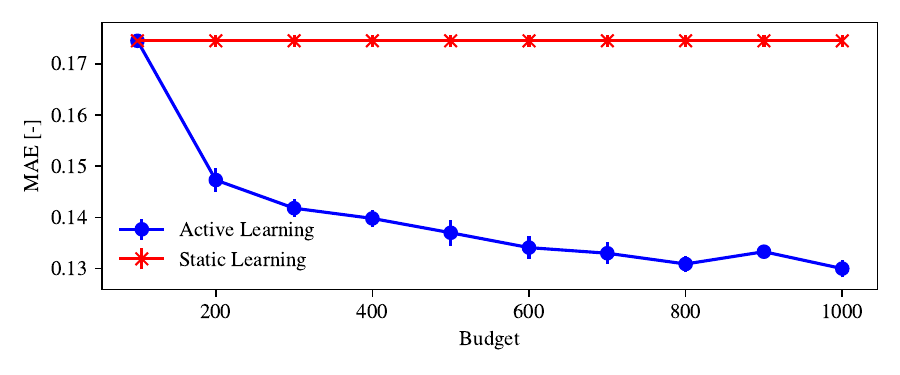} 
    \caption{Mean Absolute Error (MAE) of the models used in Active Learning (AL) and Static Learning (SL) on a hold-out test set, as a function of iterations (i.e., search budget).}

  \label{fig:ALSL-fig3}
\end{figure}

Finally, Figure~\ref{fig:ALSL-fig3} shows the Mean Absolute Error (MAE) on a hold-out test set as a function of the AL iterations (i.e., search budget). As expected, it demonstrates that in AL, the model progressively improves with each search iteration. In contrast, for SL, the accuracy remains constant, as the model does not update during the search process.

\end{document}